\documentclass[prl,twocolumn,floats,showpacs,superscriptaddress]{revtex4}
\usepackage{psfig}
\usepackage{amssymb,amsthm}
\usepackage{graphicx}
\usepackage{psfrag}
\usepackage{bm}

\newcommand{\kin}{k_{\rm{in}}}
\newcommand{\kout}{k_{\rm{out}}}

\begin{document}

\title{Detecting Network Communities: 
a new systematic and efficient algorithm}

\author{Luca Donetti}
\author{Miguel A. Mu{\~n}oz} 

\affiliation{
Departamento de Electromagnetismo y F{\'\i}sica de la Materia \\ and \\ 
Instituto de F{\'\i}sica Te{\'o}rica y Computacional Carlos I,
Facultad de Ciencias, Univ. de Granada, 18071 Granada, Spain}

\date{\today}

\begin{abstract} 
  An efficient and relatively fast algorithm for the detection of
  communities in complex networks is introduced.  The
  method exploits spectral properties of the graph Laplacian matrix
  combined with hierarchical-clustering techniques, and includes a
  procedure to maximize the ``modularity'' of the output.  Its
  performance is compared with that of other existing methods, as
  applied to different well-known instances of complex networks with a
  community-structure, both computer-generated and from the real-world.
  Our results are in all the tested cases, at least as good as the
  best ones obtained with any other methods, and faster in most of the
  cases than methods providing similar-quality results. This converts
  the algorithm in a valuable computational tool for detecting and
  analyzing communities and modular structures in complex networks.

\end{abstract} 

\pacs{89.75Hc,02.60Pn,05.50.+q} 
\maketitle 

\section{Introduction}
\label{sec:intro}

The outburst of activity in the field of Complex Networks in recent years has
been rather spectacular and amazing. Networks of any thinkable (and sometimes
``unthinkable'') type, including social, biological and technological ones
have been described, and their topological as well as dynamical features
studied. A whole line of research has emerged and a new perspective to
tackle complex problems created.  See
\cite{Strogatz,Laszlo,Porto,AleRomu,Newman} for reviews from different
perspectives and for exhaustive lists of references.

One particular aspect, which has drawn much attention, is the existence of
subsets of nodes highly linked among themselves but loosely connected to the
rest of the network, {\it i.e. communities}.  These are believed to play a
central role in the functional properties of complex structures
\cite{newman1,marta}. Identifying communities and analyzing their nature is an
important task in some fields as, for instance, computer science
\cite{Maslov,jenny}, sociology \cite{newman1,barsa}, biochemistry \cite{bio},
bibliometrics \cite{biblio}, taxonomy, or, as a more specific instance, in the
development of efficient search-engines for the WWW. According to Flake {\it
  et al.}  \cite{Flake}, ``as the web is self-organized into communities,
search-engines implementing such a concept, would help surfers to find what
they look for and avoid other contents''.

The concept of ``community'' may be retained as rather vague and
phenomenological. Indeed, depending on the network under scrutiny, it might be
quite an artificial one, while, in other cases, it emerges as a very natural
and useful structure-analysis tool. A way to make the concept more clear-cut
and practical is through the definition of the {\it modularity}, $Q$ (see
below and \cite{newman2,newmanfast}), a quantity which provides a way to
quantify the community-structure of a given network.  Other quantities have
been proposed with the same purpose \cite{Wagner,marta,roma1}.

The problem of finding communities is not new and is closely related to the
problem of graph-partitioning, profusely studied in the context of computer
science \cite{Clustering,Domany}. A review of some used techniques, including
further references, can be found in \cite{newman1,newmanepjb}. Related
problems are image processing and pattern recognition, or more generically
data-clustering: in these cases there is no underlying network, but instead
some relation or similarity between existing elements can be estabilished
\cite{image,Pattern1,Pattern2}.

In recent years many algorithms for detecting communities have been proposed,
starting with the seminal work by Girvan and Newman \cite{newman1,newman2}.
These authors proposed an iterative, {\it divisive} (as opposed to {\it
  agglomerative}) method based on the progressive removal of links with the
largest {\it betweenness}, a quantity proportional to the number of shortest
paths passing through a given edge \cite{Freeman}. The edges (or links) with
the largest betweenness have the most prominent role in connecting different
parts of the graph and, therefore, by removing them recursively a good
separation of the network into its components or communities can be found.
This method generates very good results and has been employed by different
authors in studies of various kinds \cite{Wilkinson}.  Unluckily, as already
pointed out by the authors themselves, it has a main disadvantage: its
computational demand is very high.  For instance, for sparse networks with $N$
nodes, the computation-time grows like $N^3$.  In order to deal with large
networks, for which the previous algorithm turns out to be not viable, Newman
himself developed a faster method (of the order $N^2$). It is based on the
iterative agglomeration of small communities, starting from isolated nodes, by
locally optimizing the modularity.  This method generates worse results
\cite{worse} than the previous one.

Some alternative algorithms both divisive and agglomerative (which we
do not attempt to exhaustively overview here) have been proposed in
the last months.  Some of them are listed here in chronological order
(see \cite{newmanepjb} for a more critical discussion of some of
them):

\begin{itemize}
  
\item The method by Radicchi-Castellano-Cecconi-Loreto-Parisi \cite{roma1} is
  of order $N^2$. It is a divisive algorithm that works nicely whenever
  triangular (or higher order) loops are present in the network.
  
\item Wu-Huberman algorithm \cite{huberwu}. It is a fast method
  (linear in $N$), based on the idea of voltage drops,
  which visualizes the network as an electric circuit.  It can
  be used to locate the community to which one specific node belongs,
   but it requires successive
  iterations of the method in order to provide a global network
  division in communities.

  \item Reichardt-Bornholdt method \cite{RB}. In this recent paper the
  authors introduce an algorithm inspired in the celebrated {\it
  super-paramagnetic clustering} algorithm devised by Blatt, Wiseman
  and Domany \cite{BWD}.  It is based on a $q$-state Potts Hamiltonian,
  and allows, for the first time, for the identification of fuzzy
  communities.

\item Capocci-Servedio-Colaiori-Caldarelli method \cite{roma2}.  This
  algorithm combines the use of spectral properties (which are nicely reviewed
  and generalized to study different types of networks as, for instance,
  directed ones) with the use of correlation measurements to determine
  community closeness.

\item Fortunato-Latora-Marchiori method \cite{fortunato}.  This is a variation
  of the method by Girvan and Newman, in which the betweenness is
  substituted by the alternative concept of {\it information
  centrality}, as a way to measure  edge-centrality. The method
  generates good results but its performance ($N^4$ for a sparse
  graph) is rather poor.
\end{itemize}

Apart from these techniques recently introduced in the field of complex
networks, many other algorithms have been developed mainly in the context of
computer science.  Most of them employ spectral analysis, which provides, in a
very natural way (using the first non-trivial eigenmode) a tool for
bi-partitioning \cite{specbis} as will be illustrated along this paper.  By
iterative applications of bi-partitioning more elaborated divisions into
communities or components can be achieved \cite{CS1,CS2,jenny}.
Alternatively, some other spectral methods employ more than one eigenmode
leading directly to a splitting \cite{Wagner,Kleinberg,Jordan}.

Without neglecting any of these algorithms, which can be applicable
depending on the situation under consideration, this paper introduces
yet a new method, allowing for a systematic analysis and detection of
communities. It combines the following features: {\bf i}) the
generation of good results in all the tested cases, {\bf ii}) it is
relatively fast, as compared with methods providing comparable
results, {\bf iii}) it includes a way to optimize the output, as will
be explained in what follows.

The method proposed in this paper combines spectral methods with clustering
techniques, and uses the concept of modularity in order to develop a working
algorithm.  More precisely, the main lines of the algorithm are as follows:
spectral analysis of the Laplacian matrix allows us to project the
network-nodes into an {\it eigenvector-space} of variable (tunable)
dimensionality. Afterwards, a {\it metric} is introduced in various possible
fashions, and then, finally, by applying standard clustering techniques a {\it
  dendrogram} \cite{newman1} is built up.  The modularity of possible
groupings (sections of the dendrogram) is maximized for every considered
dimension of the eigenvector-space and finally, the global maximum over all
possible number of eigenvectors ({\it i.e} dimensions of the space) is found.

In the forthcoming sections we review some basic ideas and definitions of
spectral analysis and we introduce our algorithm step by step. Then we apply
it to different workbench networks, comparing its performance with that of
other existing methods and, finally, the conclusions are presented.
 
\section{Using the Laplacian eigenvectors to detect communities}
\label{sec:method}

\subsection{Spectral analysis: Laplacian eigenvectors}

The topology of a network with $N$ vertices can be expressed through a
symmetric $N \times N$ matrix $\boldsymbol{L}$, the \emph{Laplacian} matrix
\cite{Biggs}.  The diagonal elements $L_{ii}$ are given by the degree $k_i$ of
the corresponding vertex $i$, while off-diagonal elements $L_{ij}$ are
equal to $-1$ if an edge between the corresponding vertices $i$ and
$j$ exists and $0$ otherwise. The sum of elements over every fixed row
or column is, trivially, equal to zero.  Therefore, a ``constant vector''
(with all its components taking the same value)
is an eigenvector with eigenvalue $0$.  Furthermore, since the
quadratic form
\[ \sum_{i,j=1}^n L_{ij} x_i x_j \] 
can be written as
\[ \sum_{\rm{links}} (x_i  - x_j)^2, \]
which is positive semidefinite, the eigenvalues of $\boldsymbol{L}$
are either zero or positive \cite{Mohar}.  The use of other matrices,
employed to study network spectral properties has been recently
considered in \cite{roma2,CS2,Wagner}. 


If the graph under analysis is connected, there is only one zero
eigenvalue corresponding to a constant eigenvector. On the contrary,
for non-connected graphs (composed by $m$ connected components) the
Laplacian matrix is block diagonal.  Each block is the Laplacian of a
subgraph and it admits a constant eigenvector with eigenvalue
$0$. Therefore, the Laplacian of the whole graph has $m$ degenerate
eigenvectors (corresponding to eigenvalue zero), each of them having
nonzero constant components for nodes in the associated subgraph and
$0$ in the rest.

If the subgraphs are not fully disconnected but, instead, a few links exist
between them, the degeneration disappears. This leaves only one trivial
eigenvector with eigenvalue $0$ and $m-1$ approximate linear combinations of
the old ones with slightly non-vanishing eigenvalues \cite{newmanepjb,roma2}.
As the Laplacian matrix is real-symmetric, with orthogonal eigenvectors, and
since the first of them has equal components, all the other ones must have
components whose total sum vanishes. In order to illustrate how these ideas
can be applied to identify communities, let us take, as a particular example,
the number of subgraphs to be $2$. In this case, the components of the second
(first nontrivial) eigenvector are positive for one subgraph and have to be
negative for the other, providing a clear-cut criterion to bisect the graph
\cite{specbis}.  If the two subgraphs are not very well separated, then this
distinction between positive and negative values becomes fuzzier. In such
cases, more elaborated criteria to decide how to separate into two subgroups
have been profusely studied in the specialized literature. Some of them
optimize purposely defined quantities as the {\it normalized cut} \cite{CS2}
or the {\it conductance} \cite{CS1}, which are defined as functions of the
number of links that exist between the two components and their sizes
\cite{NP}.  By iterating successive bisections, techniques to obtain more
elaborate splittings can be constructed \cite{jenny,CS2,CS1}.

An alternative strategy is to assume that if there are more
than two weakly-connected blocks it should be somehow possible to find
them all by inspecting the eigenvalue spectrum more accurately, instead 
of considering just the first non-trivial eigenmode \cite{Jordan,Kleinberg}.
Let us explore this idea, which is the one we will exploit, in more detail.
\begin{figure}[tbp]
  \centering \includegraphics[width=.9 \linewidth]{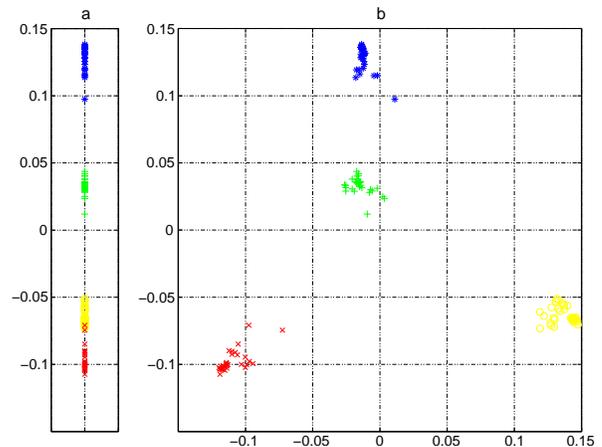} \caption{(a)
  Components of the first non-trivial eigenvector for a
  computer-generated network with $4$ communities (see main text).
  Two communities are clearly identified while the other two
  overlap. (b) All communities can be clearly identified when the
  components of the second eigenvector are plotted versus those of the
  first one; {\it i.e.} when the dimensionality of the
  eigenvector-space is enlarged.}  \label{fig:ev}
\end{figure}
Figure~\ref{fig:ev}a shows the components of the first nontrivial eigenvector
of a computer-generated graph including 4 communities, each composed by 32
nodes (see forthcoming sections for details). The group structure is clear,
even if the two communities at the bottom are very near to each other and some
nodes could be mis-classified. In other examples, with a number of inter-group
connections larger than here, the communities become more entangled, and the
prospective of extracting clear-cut subdivisions using this type of
one-dimensional plot worsens. This difficulty can be circumvented by taking
into account some more eigenvectors, {\it i.e.}  by enlarging the
projection-space.  This is illustrated in figure~\ref{fig:ev}b, where the
nodes of the same graph are plotted using the components on the first two
nontrivial eigenvectors as coordinates.  Simple eye-inspection shows that all
communities are distinctly separated now.  Actually, using three eigenvectors
the nodes of the different groups fall around the vertices of a (slightly
distorted) tetrahedron, with some further improvement in inter-community
separation. Generalizing this idea, {\it each vertex in the graph is
  represented by a point in a $D$-dimensional space in which the coordinates
  are given by its projections on the first $D$ nontrivial eigenvectors}.

\subsection{Introducing a metric}

Aimed at turning ``eye-inspection'' of communities into a more quantitative
measure, the explicit introduction of a metric (or similarity measure) is
required. The most straightforward choice would be the Euclidean distance.
However, this is {\it not} the only possibility; another one is to consider
the {\it angular distance}, defined as the angle between the vectors joining
the origin of the $D$-dimensional space with the two points under
consideration. This possibility is inspired by empirical observations: loosely
connected nodes could be quite ``Euclideanly'' far from each other within a
community, but still lying in the same ``direction'' in the eigenvector-space
\cite{cc}.  Moreover, when networks are large, nodes in the same community
form a roughly one-dimensional ``bundle'' (see for example figure 3 in
\cite{Maslov}).  Note also, that using angular distances is tantamount to
normalizing the position-vectors in the corresponding space and then measuring
the Euclidean distance, similarly to what proposed in \cite{Jordan}. As will
be shown, the angular metric generates, as a matter of fact, better results
than the Euclidean one.

\subsection{Cluster analysis}

Having introduced a way to measure distances in the eigenvector space, a
method to group nodes into communities is required.  Such a method is
provided by standard clustering techniques \cite{Clustering} as,
for example, {\it hierarchical clustering}. Starting from $N$ clusters,
composed by individual nodes, the two closest ones are iteratively
joined together. 
In order to define cluster-to-cluster distance or ``closeness'' (for a
given metric) different criteria can be employed, generating among
others, the following clustering algorithms \cite{Clustering}:

\begin{itemize}
  
\item All possible pairs of nodes, taking one from each of the two clusters
  under examination, are considered. The minimum possible node-to-node
  distance is declared to be the cluster-to-cluster closeness.  This leads to
  {\it single-linkage clustering}.
  
\item Proceeding as before, but replacing the ``minimum possible node-to-node
  distance'' between pairs by the ``maximum'' one, {\it complete-linkage
    clustering} is defined.

\item Another possibility consists in taking the average distance between all
  possible pairs. This leads to {\it group-average clustering}.

\item A cluster is represented by a single point located at its ``center of
  mass''; the cluster-to-cluster distance is defined as the node-to-node
  distance between these two points. This leads to {\it centroid clustering}.
\end{itemize}

All these criteria have been broadly studied and applied. None of them
can be proved to be generically more efficient than the others. In
particular, the single-linkage method, being very simple, can be
useful to analyze large data sets, and possesses some further
mathematical advantages \cite{Clustering}.  On the contrary, it has a
tendency to cluster together, at a relatively low level, distant nodes
linked successively by a series of intermediates. This is usually
called {\it chaining property}, which constitutes in some cases a
serious drawback.

On the other hand, a convenient advantage of both, single and complete-linkage
clusterings, is that only the ordering of the similarity measure is important:
every other metric which produces the same ordering of distances leads to the
same results.

The output of these algorithms can be represented by a hierarchical tree
usually called \emph{dendrogram}. The starting single-node communities are the
branch-tips of such a tree, which are repeatedly joined until the whole
network has been reconstructed as a single component (see, for instance,
figure 2 in \cite{newman1}). Each level of the tree represents a possible
splitting of the network into a set of communities, obtained by halting the
clustering process at the corresponding level. However, the clustering
algorithm gives no hint about the ``goodness'' of such a partition.

\subsection{Modularity}

In order to quantify the validity of possible sub-divisions (obtained
as explained above) and to optimize the chosen splitting, we use,
following \cite{newman2,newmanfast}, the concept of
\emph{modularity}. It is defined as follows: given a network division,
let $e_{ii}$ be the fraction of edges in the network between any two
vertices in the subgroup $i$, and $a_i$ the total fraction of edges
with one vertex in group $i$ (where edges ``internal'' to each group
have weight $1$ while inter-group links are weighted $1/2$). The
modularity, $Q$, is then defined as
\begin{equation}
  \label{eq:modularity}
  Q = \sum_i (e_{ii} - a_i)^2.
\end{equation}
It measures the fraction of edges that fall between communities minus the
expected value of same quantity in a random graph with the same community
division.

The maximization of modularity has been proposed as a possible way to
detect communities; since a full maximization is not possible in
practice (the algorithm would take an amount of time exponential in
the number of nodes to explore all possible splittings) an
approximate algorithm has been suggested \cite{newmanfast}. In our
case, modularity measurements are simply used to find the best
splitting among all the possible partitions of the dendrogram obtained
following the previous steps \cite{newman2}.

Other indeces quantifying the quality of splittings have been also proposed in
the literature. Some of them are the ``conductance'', the ``performance'', and
the ``coverage'' to name but a few (see \cite{Wagner} and references therein
for more details). None of these taken by itself, provides a fully useful
criterion; they have to be combined somehow. It seems that the modularity is
a better, more efficient, choice.

\subsection{Implementing a functioning algorithm}

Summarizing the ideas introduced in the previous sub-sections, our algorithm
can be synthesized and implemented to build up a functioning algorithm as
follows.  First {\it a few} eigenvalues and eigenvectors of the network
Laplacian matrix are computed. The question of what ``a few'' means will be
tackled afterwards.  Since the Laplacian is usually a sparse matrix and not
all eigenpairs are required (that will require a time $N^3$) the relatively
fast Lanczos method \cite{lanczos} can be employed.  Nonetheless, the
eigenvector computation is still the most computationally expensive step of
the algorithm.

For any given number $D$ of eigenvectors ({\it i.e.} for a fixed dimension of
the space) a similarity measure (or metric) is chosen, providing a basis to
apply one of the previously introduced clustering techniques. Typically,
Euclidean or, better, angular distances are employed.

Among the various hierarchical clustering methods available, we test single-
and complete-linkage clustering algorithms. These two have the advantage that
no new distances have to be calculated during cluster formation: when two
subgroups merge to form a larger one, its distance to any other cluster is
given by the shortest (single-linkage) or by the largest (complete-linkage) of
the distances from the two original components. As said before, single-linkage
performs poorly in many cases owing to the previously discussed ``chaining''
property, converting complete-linkage in the preferential choice. Other
linkage methods will be explored in the future; in particular, group-average
linkage could be suitable when studying tree-like graphs \cite{private}.

An important difference between the way we apply clustering techniques and
other standard applications is that we know in advance the underlying network
structure. Using this knowledge we implement the constraint that {\it two
  clusters are susceptible to be merged only if there exists a link between
  them in the original network}.

At every step of the clustering process the modularity is computed.  Once the
whole dendrogram is completed, the splitting with the maximum modularity is
chosen as the output for the corresponding $D$.

The optimal value of $D$ to be taken is not known \textsl{a priori}, but since
the eigenvalue calculation is the slowest part of the algorithm, we can repeat
the hierarchical clustering using all possible values of $D$, and look up for
the largest value of the modularity.  Typically the largest-modularity vs. $D$
curve exhibits a maximum whose corresponding splitting provides the algorithm
final output. If, instead, the curve keeps on growing up to the largest $D$,
the number of computed eigenpairs has to be enlarged, in order to extend the
range of the curve, until a clear-cut maximum is pin-pointed.

\section{Tests of the method}
\label{sec:test}

\subsection{Artificial community networks}
\label{sec:artif}

To prove the algorithm we first test it on computer-generated random
graphs with a well-known pre-determined community structure
\cite{newman1}. Each graph has $N=128$ nodes divided into $4$
communities of $32$ nodes each. Edges between two nodes are
introduced with different probabilities depending on whether the two
nodes belong to the same group or not: every node
has $\kin$ links on average to its fellows in the same community, and
$\kout$ links to the outer-world, keeping  $\kin+\kout=16$.

\begin{figure}[tbp]
  \centering \includegraphics[width=.95 \linewidth]{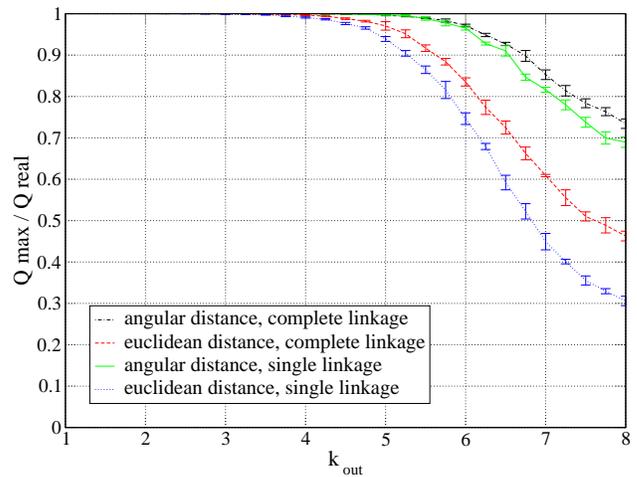}
  \caption{Maximum modularity found by the algorithm, divided by that
  of the known splitting of a computer generated random graph (see
  main text); the average over $200$ graphs is plotted as a function of
  $\kout$.}  \label{fig:rc1}
\end{figure}
\begin{figure}[tbp]
  \centering
  \includegraphics[width=.95 \linewidth]{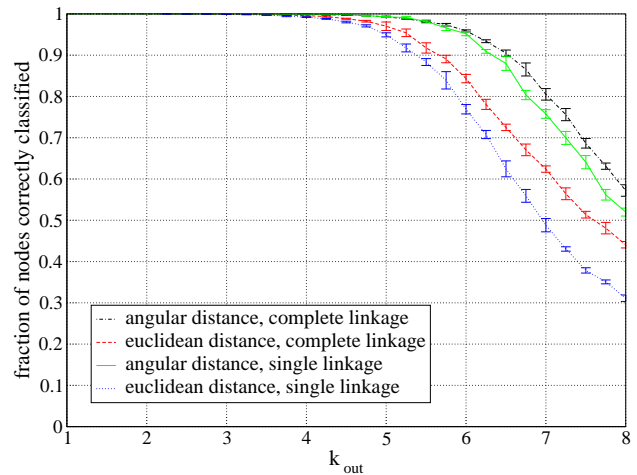}
  \caption{Fraction of nodes of computer generated random graphs correctly 
    identified by the algorithm, averaged over $200$ graphs, as a function 
    of $\kout$.}
  \label{fig:rc2}
\end{figure}

In figures~\ref{fig:rc1} and \ref{fig:rc2} we plot the modularity
corresponding to the best splitting identified by the algorithm normalized by
the one of the known answer, and the average number of correctly classified
vertices, respectively.  Data for both, Euclidean and angular measures, and
both, single- and complete-linkage algorithms, are shown. The number of
eigenvalues leading to the largest modularity is between $3$ and $5$ for the
angular distance, and between $2$ and $4$ for the Euclidean one. Let us remark
that these are roughly equal to the number of communities and that the
performance is much better using the angular distance.

Summing up: on these computer-generated networks, our algorithm (equipped with
the angular distance and complete-linkage) generates excellent results as
compared with other methods (see, for instance, figure~1 in \cite{newmanfast}
and figure~3 in \cite{fortunato}).

\subsection{Zachary karate club}

Now we consider the well-known karate club friendship network studied by
Zachary \cite{zachary}, which has become a commonly used workbench for
community-finding algorithms testing
\cite{roma1,huberwu,newman1,newman2,newmanfast,RB,fortunato}.

\begin{table}
  \centering
  \begin{tabular}{|r||r|r|}
    \hline
    & angular & Euclidean \\
    \hline\hline
    single-linkage & 0.412 & 0.319 \\
    complete-linkage & 0.412 & 0.368 \\
    \hline
  \end{tabular}
  \caption{Modularity of the best splitting of the Zachary club
    network obtained for different metrics and
    clustering algorithms.}
  \label{t:karate}
\end{table}

Table~\ref{t:karate} shows the maximum modularity found by the algorithm: the
best value is again obtained using angular distances combined with either
single or complete-linkage clustering.
\begin{figure}[tbp]
  \centering
  \includegraphics[width=.9 \linewidth]{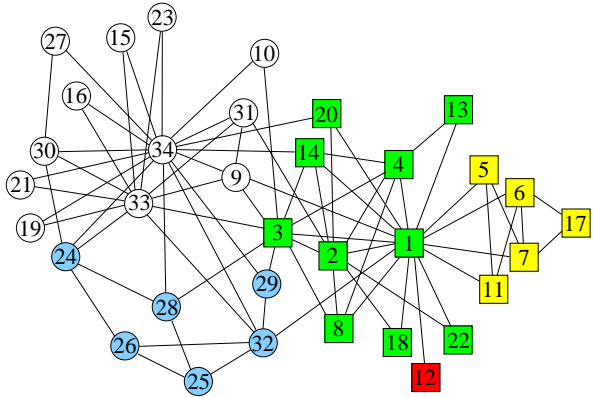}
  \caption{Splitting of the Zachary club network. Squares and circles
    indicate the two communities observed by Zachary, colors denote the further
    subdivision found by our algorithm.}
  \label{fig:zac}
\end{figure}

The best splitting is shown in figure~\ref{fig:zac}; it is different from the
``actual'' breakdown of the club; {\it i.e.} the two groups reported by
Zachary are further subdivided. Let us stress the presence of a single-node
community (node $12$), and the fact that the modularity value of this
splitting is larger than Zachary's one ($0.371$), and larger that the ones
found using other methods \cite{newmanfast,RB,fortunato}.


\begin{figure}[tbp]
  \centering
  \includegraphics[width=.9 \linewidth]{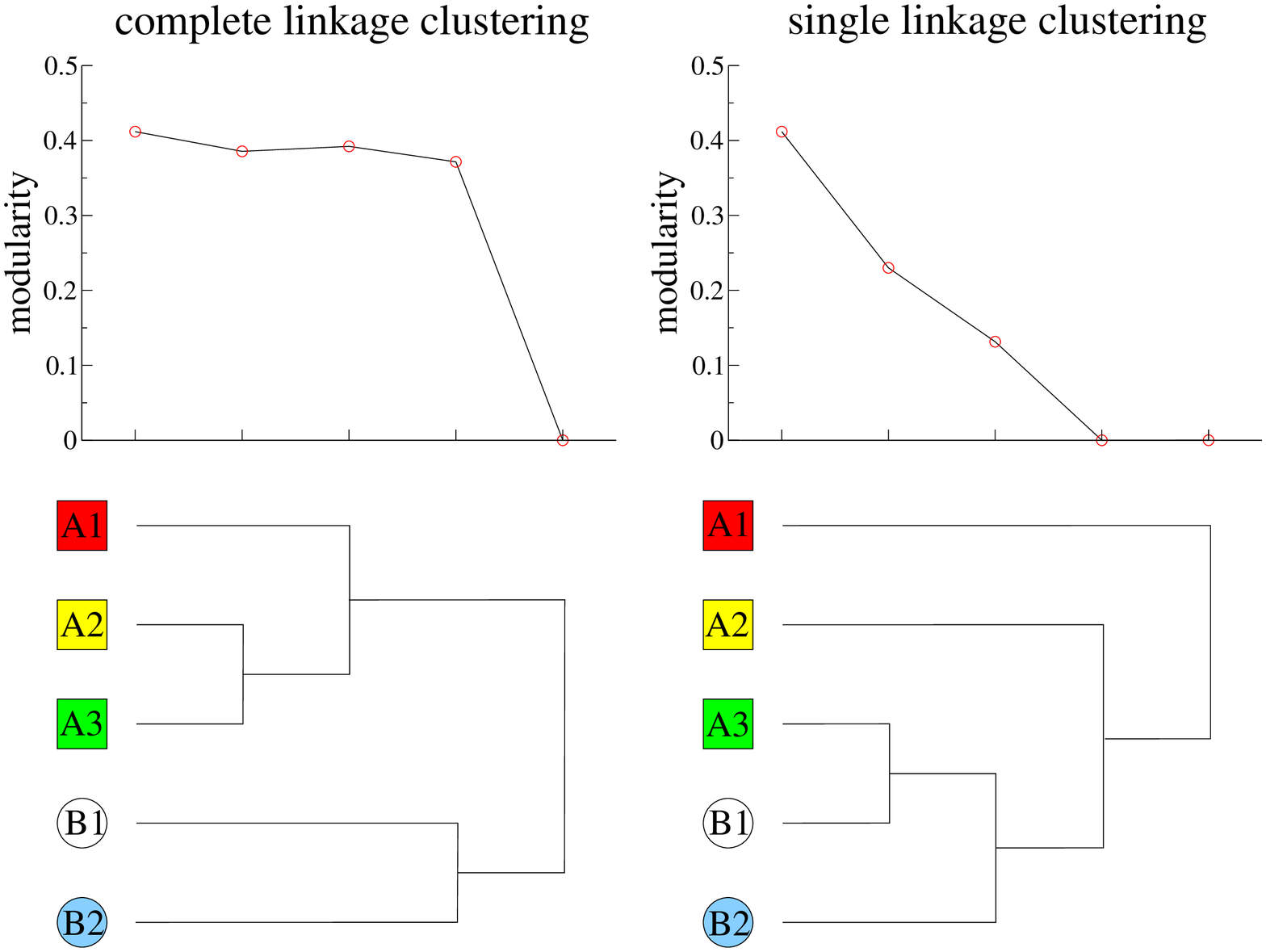}
 \caption{Comparison between the final part of the dendrogram for 
the Zachary club, by using complete- and single-linkage clustering (bottom), 
together with the corresponding modularity values (top).}
  \label{fig:den}
\end{figure}

In this case single and complete-linkage give the same best splitting.
Nevertheless, the hierarchical structure given by the dendrogram in the two
cases are quite different.  Figure~\ref{fig:den} shows how clusters merge
after the best splitting is identified, as well as the modularity value
corresponding to each division.  For complete-linkage the modularity value
remains close to the best one until the whole network is merged in one
community.  On the other hand, for single-linkage it falls down rather
abruptly right after the first merging, owing to the chaining problem.
Moreover, in the former case, the two Zachary communities are first
reconstructed and then joined together, while in the latter the merging
proceeds differently. Therefore, even if the best splitting is the same one in
both cases, complete-linkage produces a more reliable dendrogram, describing
more accurately the hierarchical structure.

\subsection{Scientific collaboration networks}

In order to test the method performance on larger networks we
consider two scientific-collaboration networks first analyzed by
Newman \cite{newman-data}. The vertices are the authors of the papers
appeared in the \texttt{cond-mat} and \texttt{hep-th} archives at
\texttt{ArXiv.org} between $1995$ and $1999$. Two authors are linked
if they have co-authored a paper together.

\begin{figure}[tbp]
  \centering
  \includegraphics[width=.9 \linewidth]{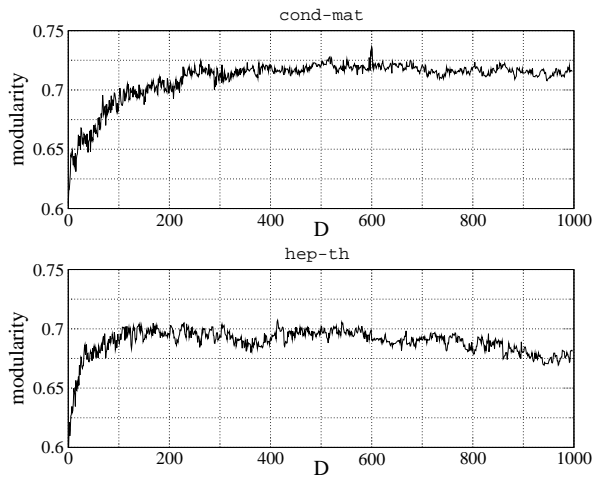}
  \caption{Maximum modularity as a function of the number  of eigenvectors 
    for the \texttt{cond-mat} (top) and \texttt{hep-th} (bottom) networks.}
  \label{fig:qmaxcm}
\end{figure}

The \texttt{cond-mat} network contains $16726$ nodes, but we focus on its
largest connected component, which contains only $13861$ authors. The
computation of the first $1000$ eigenvectors takes about two hours on a
personal computer. The modularity curve computation, calculated using up to
$999$ eigenvectors, lasts around 15 minutes.  Results for angular distance and
complete-linkage clustering are plotted in figure~\ref{fig:qmaxcm}.
The largest value of the modularity, $Q=0.736$, achieved for a splitting in
$229$ communities, corresponds to a $602$-dimensional space.  Obviously, we
cannot compare the final splitting with a ``true'' one, which is not defined.
As the curve in figure \ref{fig:qmaxcm} is rather flat in its tail, one can
legitimately wonder how does the best splitting compare to other ones obtained
for similar dimensionalities.  This question is difficult to answer in a
rigorous way, and will deserve further analysis, which will eventually lead to
a functional definition of the {\it community-structure robustness}.

Analogously, the \texttt{hep-th} network has $8361$ authors with a connected
component of $5835$.  The largest modularity value, $Q = 0.707$, is produced
by a division into $114$ communities, obtained using $416$ eigenvectors. The
computation of the first $1000$ eigenvectors takes around $30$ minutes and the
search of the largest modularity value about $8$ minutes.  In this case,
the initial number of eigenvectors could have been taken much smaller than
$1000$, without affecting the final output, with the consequent time saving.

As in previous cases, the number of eigenvectors used to produce the best
splitting is of the order of magnitude of the number of found communities.  In
these cases, comparison with previous community studies is not feasible, as
modularity measurements have not been (to the best of our knowledge) reported
in literature.

\section{Conclusions}
\label{sec:concl}

We have introduced a new algorithm aimed at detecting community structure in
complex networks in an efficient and systematic way. The method combines
spectral techniques, cluster analysis, and the recently introduced concept of
modularity.

The nodes of the network are projected into a $D$-dimensional space, where $D$
is a number of first non-trivial eigenvectors of the Laplacian matrix; their
coordinates are the node-projections on each eigenvector. Then a metric
(either Euclidean or angular) is introduced in such an eigenvector space.
Once distances are computed, standard hierarchical clustering techniques (as,
for instance, complete-linkage clustering) are employed to generate a
dendrogram. The subdivision of this dendrogram giving the maximum modularity
is taken as the output of the algorithm for a fixed $D$. Then, also $D$ is
allowed to vary (from $1$ to some arbitrary, maximum value) providing a way to
maximize the modularity and enhance the performance of the method.

The best results are obtained using the angular distance and complete-linkage
clustering; however, other types of distances, other clustering algorithms, or
even other means to quantify the goodness of a division could be used to
improve the results. In this sense our algorithm is a ``block-modular'' one:
modifications of any of its ingredients could possibly lead to an overall
improvement.

Even if spectral methods have been profusely used before to analyze similar
problems, we believe that our algorithm represents a step forward in studying
complex-network communities, as it combines spectral techniques with (i) the
novel concept of modularity, which provides a very adequate estimate of the
quality of a given splitting and (ii) a way to optimize the number of
eigenmodes taken into consideration.

The weakest part of the method is that the maximum number of eigenvectors to
be computed in order to find the one generating the maximum modularity is not
known a priori. Being the calculation of eigenvectors the slowest part of the
algorithm, what we do is to take a reasonable number of them and, afterwards,
verify that the maximum-modularity curve as a function of $D$ decreases at its
tail; {\it i.e.} we make sure that a maximum of the modularity function is
located. If this is not the case, the number of eigenvectors needs to be
enlarged, at the cost of higher computational effort. In the absence of a
general criterion to establish the monotonicity of the modularity curve, the
only possible way to decide whether the identified local maximum is the global
one, would be to compute all possible eigenvalues. In practice, in all the
studied cases, the best splitting is found with a relatively small number of
eigenvectors, converting the algorithm in a reliable, relatively fast, and
very efficient one.

An open challenge would be identifying a systematic criterion to
estimate, a priori, what is the order of magnitude of the number of
eigenvalues to be computed to further optimize the output and
efficiency.

We hope that this new algorithm will be employed with success in the search
and study of communities in complex networks, and will help to uncover new
interesting properties.

\begin{acknowledgments}
  We acknowledge useful comments and discussions with F. Colaiori, A. Capocci,
  V. Servedio, A. Arenas, G. Caldarelli, and J. Torres.  We are specially
  grateful to M. Newman for providing us with the data on scientific
  collaborations as well as for a reading of the manuscript, and to C.
  Castellano, for very helpful comments and suggestions.  Financial support
  from the Spanish MCyT (FEDER) under project BFM2001-2841 and the EU COSIN
  project IST2001-33555 is acknowledged.
\end{acknowledgments}


\begin{thebibliography}{99}
  
\bibitem{Strogatz} S. H. Strogatz, Nature {\bf 410}, 268 (2001).

\bibitem{Laszlo} A. L. Barab\'asi,  
Rev. Mod. Phys. {\bf 74}, 47 (2002).

\bibitem{Porto} S. N. Dorogovtsev and J. F. F. Mendes, {\it Evolution of
    Networks: From Biological Nets to the Internet and WWW}, Oxford
  University Press, Oxford (2003).
  
\bibitem{AleRomu} R. Pastor Satorras and A. Vespignani, {\it Evolution and
    Structure of the Internet: A Statistical Physics approach}, Cambridge
  University Press (2004).
  
\bibitem{Newman} M. E. J. Newman,
SIAM Review {\bf 45}, 167 (2003).

\bibitem{newman1} M. Girvan, M. E. J. Newman, Proc. Natl. Acad. Sci. USA {\bf
    99}, 7821-7826 (2002).

\bibitem{marta}  R. Guimer\'a, M. Sales-Pardo, and L. A. N. Amaral,
cond-mat/0403660.

\bibitem{Maslov} K. A. Eriksen, I. Simonsen, S. Maslov, and K. Sneppen,
Phys. Rev. Lett. {\bf 90}, 148701 (2003); and cond-mat/0312476.

\bibitem{jenny} C. Borgs, J. T. Chayes, M. Mahdian and A. Saberi, Proceedings
  of the 10th ACM SIGKDD International Conference on Knowledge, Discovery and
  Data Mining (2004).

\bibitem{barsa} R. Guimer\'a, L. Danon, A. Diaz-Guilera, F. Giralt, and A.
Arenas, Phys. Rev. E {\bf 68}, R065103 (2003); A. Arenas,  L. Danon, 
A. Diaz-Guilera, P. M. Gleiser, and  R. Guimer\'a, cond-mat/0312040.

\bibitem{bio} L. H. Hartwell, J. J. Hopfield, S. Leibler, and A. W. Murray,
  Nature {\bf 402}, C47 (1999).  E. Ravasz, A. L. Somera, D. A. Mongru, Z. N.
  Olvai, and A. L. Barab\'asi, Science {\bf 297}, 1551 (2002).

\bibitem{biblio} L. Egghe and R. Rousseau, {\it Introduction to Informetrics},
(1990). 

\bibitem{Flake} G. W. Flake, S. Lawrence, C. L. Giles, and F. Coetzee, IEEE
  Computer {\bf 35}, 66 (2002).

  
\bibitem{newman2} M. E. J. Newman, M. Girvan, Phys. Rev. E {\bf 69}, 026113
  (2004).
  
\bibitem{newmanfast} M. E. J. Newman, Phys. Rev. E {\bf 69}, 066133 (2004)
  
\bibitem{Wagner} U. Brandes, M. Gaertler, and D. Wagner.  Proc. 11th Europ.
  Symp. Algorithms (ESA'03), LNCS 2832, pp. 568-579.
  
\bibitem{roma1} F. Radicchi, C. Castellano, F. Cecconi, V. Loreto, and D.
  Parisi, Proc. Natl. Acad. Sci. USA {\bf 101}, 2658-2663 (2004).
  
\bibitem{Clustering} A. K. Jain and R. C. Dubes, {\it Algorithms for
    clustering data}, Prentice Hall, Englewood Cliffs, NJ (1988).  B. S.
  Everitt, {\it Cluster Analysis}, Edward Arnold, London (1993).

\bibitem{Domany} E. Domany, Physica A {\bf 263}, 158 (1999).
  
\bibitem{newmanepjb} M. E. J. Newman, Eur. Phys. J. B {\bf 38}, 321-330
  (2004).
  
\bibitem{image} Y. Weiss, Proceedings IEEE International Conference on Computer
    Vision, 975-982 (1999).
  
\bibitem{Pattern1} R. O. Duda and P. E. Hart, {\it Pattern Classification and
    Scene Analysis}, Wiley. New York (1973).
  
\bibitem{Pattern2} K. Fukunaga, {\it Introduction to statistical pattern
    recognition}, Academic Press, San Diego (1990).
  
\bibitem{Freeman} L. Freeman, Sociometry {\bf 40}, 35 (1977).
   
\bibitem{Wilkinson} J. R. Tyler, D. M. Wilkinson, and B. A. Huberman, in {\it
    Proceedings of the First International Conference on Communities and
    Technologies}, Ed. M. Huysman, E. Wenger, and V. Wulf, Kluwer, Dordrecht
  (2003). D. Wilkinson and B. A. Huberman, Proc. Natl. Acad. Sci. USA {\bf
    101}, 5241-5248 (2004).
    
\bibitem{worse} The goodness of a given division (or division method) can be
  decided in {\it absolute} terms (when the underlying
  community-structure is known, as for example, in computer-generated
  networks) or in {\it relative} terms (when the community structure
  is not known, but it maybe quantified in terms of modularity or
  similar measurements \cite{newman2,newmanfast,roma1}; large
  modularity-values corresponding to better divisions).
   
\bibitem{huberwu} F. Wu, B. A. Huberman, Eur. Phys. J. B {\bf 38}, 331-338
  (2004).
  
\bibitem{RB} J. Reichardt, S. Bornholdt, cond-mat/0402349.
  
\bibitem{BWD} M. Blatt, S. Wiseman, and E. Domany, Phys. Rev. Lett.  {\bf 76},
  3251 (1996); Neural Computation {\bf 9}, 1805 (1997).

\bibitem{roma2} A. Capocci, V. Servedio, F. Colaiori, and G. Caldarelli,
  cond-mat/0402499.
  
\bibitem{fortunato} S. Fortunato, V. Latora, M. Marchiori, cond-mat/0402522.
 
\bibitem{specbis}
  M. Fiedler, Czech. Math. J. {\bf 23}, 298-305 (1973) 
  A. Pothen, H. Simon, K.-P. Liou, SIAM J. Matrix Anal. Appl. {\bf 11},
  430-452, (1990)
 
\bibitem{CS1} R. Kannan, S. Vempala, and A. Vetta, Journal of the ACM {\bf
    51}, 497-515 (2004)

\bibitem{CS2} X. He, C. H. Q. Ding, H. Zha, and H. D. Simon, {\it Proc. IEEE
    Int'l Conf. Data Mining}, 195 (2001).  C. H. Q. Ding, X. He and H. Zha.
  {\it Proccedings 7th International Conf. on Knowledge Discovery and Data
    Mining (KDD 2001)}, 275 (2001).

\bibitem{Kleinberg} J. M. Kleinberg. 
 Journal of the ACM {\bf 46}, 604 (1999). 
 D. Gibson, J. M. Kleinberg, and P. Raghavan. 
{\it Proceedings of the 9th ACM 
 Conference on Hypertext and Hypermedia}, 225 (1998).

\bibitem{Jordan} A. Y. Ng, M. I. Jordan, and Y. Weiss, Advances in Neural
  Information Processing Systems {\bf 14}, 849 (2002).

\bibitem{Biggs} N. L. Biggs, \emph{Algebraic Graph Theory}, Cambridge
  University Press (1974).
  
\bibitem{Mohar} B. Mohar, The Laplacian spectrum of graphs, in: Y. Alavi, G.
  Chartrand, O. R. Ollermann, A.J. Schwenk (Eds.),{\it Graph Theory,
    Combinatorics, and Applications}, Wiley, New York, 1991, pp. 871-898.
  
\bibitem{NP} In principle the minimization of the conductance or the
  normalized-cut among all possible splits is a NP-hard problem. However, it
  can be shown that the cuts based on the components of the second eigenvector
  of the Laplacian or some related matrix give a guaranteed approximation to
  the optimal cut \cite{Jordan,Chung}.

\bibitem{Chung} F. Chung. {\it Spectral Graph Theory}, Number 92 in CBMS
  Region Conference Seriesin Mathematics. American Mathematical Society,
1997.
  
\bibitem{cc} The attentive reader could argue that figure 1b provides a
  counterexample to this general assertion; {\it i.e.} the two uppermost
  groups are nearby angularly but far apart Euclideanly. Indeed, taking the
  three-dimensional version of the net analyzed in such a figure, the four
  communities lay within the main directions on a tetrahedron, circumventing
  this apparent contradiction.
  
\bibitem{lanczos} G. H. Golub, C. F. Van Loan, {\it Matrix Computations},
  Johns Hopkins University Press, Baltimore (1996).

\bibitem{private} M. Newman, private communication. 
  
\bibitem{zachary} W. W. Zachary, J. of Anthropological Research {\bf 33}, 452
  (1977).
  
\bibitem{newman-data} M. E. J. Newman, The structure of scientific
  collaboration networks, Proc. Natl. Acad. Sci. USA {\bf 98}, 404-409 (2001).
  See also, M. E. J. Newman, Phys. Rev. E {\bf 64}, 016132 (2001).
  
\end{thebibliography}
\end{document}